\documentclass{article}
\usepackage{amssymb}
\usepackage{amsfonts}
\usepackage{amsmath}

\input{tcilatex}

\begin{document}

\title{Direct coupling between magnetism and superconducting current in
Josephson $\varphi _{0}$ junction}
\author{A. Buzdin \\
Institut Universitaire de France and \\
Universit\'{e} Bordeaux I, CPMOH, UMR 5798, \\
33405 Talence, France}
\maketitle

\begin{abstract}
We study the proximity effect between conventional superconductor and
magnetic normal metal with a spin-orbit interaction of the Rashba type.
Using the phenomenological Ginzburg-Landau theory and the quasiclassical
Eilenberger approach it is demonstrated that the Josephson junction with
such a metal as a weak link has a special non-sinusoidal current-phase
relation. The ground state of this junction is caracterized by the finite
phase difference $\varphi _{0},$ \ which is proportional to the strength of
the spin-orbit interaction and the exchange field in the normal metal. The
proposed mechanism of the $\varphi _{0}$ junction formation gives a direct
coupling between the superconducting current and the magnetic moment in the
weak link. Therefore the $\varphi _{0}$ junctions open interesting
perspectives for the superconducting spintronics.
\end{abstract}

\bigskip \bigskip

Usually the current-phase relations in Josephson junctions near the critical
temperature are sinusoidal $j(\varphi)=j_{c}\sin(\varphi)$ but with lowering
temperature the contribution of the higher harmonic terms $\sim
j_{n}\sin(n\varphi)$ can be observed. However if the time reversal symmetry
is preserved the current-phase relation is always antisymmetric $j(-\varphi
)=-j(\varphi)$ \cite{GolubovRMP}. Without this restriction a more general $%
j(\varphi)=j_{0}\sin(\varphi+\varphi_{0})$ dependence is also possible and
the generic expression for the current in the pioneering work of Josephson 
\cite{Josephson}\ incorporates this possibility. In fact such current-phase
relations have been predicted for Josephson coupling involving the
unconventional superconductors \cite{Geshkenbein},\cite{Yip},\cite{Sigrist},%
\cite{Tanaka}. The experimental verification of these predictions are still
lacking.

In the present work we demonstrate that the Josephson superconductor/normal
metal/superconductor junctions (S/N/S) \ provides the realization of such
unusual current-phase relations $j(\varphi)=j_{0}\sin(\varphi+\varphi_{0})$
for the case of conventional superconductors when the normal layer is a
non-centrosymmetric, i. e. with broken inversion symmetry (BIS) magnetic
metal. Further on we will call this junction " $\varphi_{0}$ junction". \
The phase shift $\varphi_{0}$ is proportional to the magnetic moment, and
therefore these $\varphi_{0}$ junctions serve example of systems with direct
coupling between magnetic moment (internal exchange field) and
superconducting current. This opens an interesting field of application of $%
\varphi_{0}$ junctions in superconducting spintronics. Varying the N layer
thickness we may easily control the phase shift $\varphi_{0}.$ Note that the
considered situation is different from the case of the Josephson junction
with dominating second sinusoidal harmonic - see \cite{Tanaka},\cite%
{Goldobin} and references cited therein. In such a case at the ground state
an arbitrary phase drop across the junction may exist if the sign of the
second harmonic is negative. However in these systems it is impossible to
have the direct coupling between magnetic exchange field and superconducting
phase and the properties of these junction are very different from that of $%
\varphi_{0}$ junctions considered here.

\bigskip Before addressing the problem of the proximity effect between
conventional superconductor and BIS magnetic metal it is useful to recall
that recently BIS superconductors attracted a lot of attention. Namely the
heavy fermion superconductor CePt$_{3}$Si provides a famous example of the
superconductivity and antiferromagnetism coexistence in the
noncentrosymmetric compound \cite{Bauer}. Now the number of superconductors
without inversion symmetry approaches one dozen and during the last years
their properties were under intense studies from both theoretical and
experimental points of view, see \cite{Fujimoto},\cite{Samokhin1},\cite%
{Samokhin2},\cite{Samokhin3},\cite{Kaur},\cite{Frigeri} and references cited
therein. In the presence of the magnetic field the lack of inversion
symmetry leads to the spatially modulated helical superconducting phase \cite%
{Samokhin1},\cite{Kaur},\cite{Edelstein},\cite{Dimitrova}.

\bigskip The Josephson junctions between conventional superconductor and BIS
superconductors should reveal some special features \cite{Kaur},\cite%
{Hayashi}. We stress that the aim of the present work is to study the very
different situation : the Josephson junction between conventional
superconductors with a weak link formed by BIS magnet. As an example of the
suitable candidates for such interlayer we may cite MnSi and FeGe. The
anomalous properties of studied junctions are related to the particularities
of the superconducting proximity effect in the BIS metal.

\bigskip On the microscopical level the special character of the electron
spectrum in BIS metal may be described by the Rashba-type spin-orbit
coupling \cite{Rashba}: $\alpha(\overrightarrow{\sigma}\times\overrightarrow{%
p})\cdot\overrightarrow{n}$, where $\overrightarrow{n}$ is the unit vector
along the asymmetric potential gradient and parameter $\alpha$ describes its
strength. To illustrate the unusual properties of the BIS Josephson junction
we start with a simple Ginzburg-Landau (GL) approach. Describing the weak
link by the GL theory we assume the temperature is above the critical
temperature of the material of the weak link and the superconducting order
parameter is induced only by the superconducting banks. As it has been noted
in \cite{Samokhin1},\cite{Kaur}, the Rashba-type interaction in the presence
of the field $\overrightarrow{h}$ acting on the electron spin leads to the
following GL free energy density

\bigskip

\begin{align}
F& =a\left\vert \psi \right\vert ^{2}+\gamma \left\vert \overrightarrow{%
\mathbf{D}}\psi \right\vert ^{2}+\frac{b}{2}\left\vert \psi \right\vert ^{4}
\notag \\
& -\varepsilon \overrightarrow{n}\cdot \left[ \overrightarrow{h}\times
\left( \psi \left( \overrightarrow{\mathbf{D}}\psi \right) ^{\ast }+\psi
^{\ast }\left( \overrightarrow{\mathbf{D}}\psi \right) \right) \right] ,
\label{GL}
\end{align}%
where $\psi $ is the superconducting order parameter, $D_{i}$ $=-i\partial
_{i}-2eA_{i\text{ }}$and the coefficient $a$ becomes zero at some
temperature $T_{c0}:$ $a$ $\sim (T-T_{c0})$. \ The special character of the
BIS superconductivity is described by the last term in (\ref{GL}) with the
coefficient $\varepsilon \sim \alpha $. In principle the field $%
\overrightarrow{h}$ may be created by the applying external field but we
suppose that BIS metal is a ferromagnet and $\overrightarrow{h}$ is an
internal exchange field (which is assumed to be small to avoid the necessity
to add higher order derivative terms in (\ref{GL}) \cite{Buzdin-rew}) . Note
that the origin of the spin-orbit contribution in (\ref{GL}) may be
intrinsic resulting from the crystal symmetry or extrinsic. The latter case
corresponds for example to the ferromagnetic layer with the in-plane
magnetization and a varying thickness. To be more specific below we consider
the intrinsic BIS metal. Schematically the Josephson junction is presented
in Fig. 1. Further on to concentrate on the special properties of this
junction we neglect the orbital effect. In Fig. 1 the magnetization is along 
$z-$axis and the demagnetization factor N=1 and then the internal magnetic
field in the junction $\overrightarrow{H}_{i}=-4\pi \overrightarrow{M}$.
Therefore the magnetic induction $\overrightarrow{B}=\overrightarrow{H}%
_{i}+4\pi \overrightarrow{M}=0$ and then for this geometry the orbital
effect is vanishing. Alternatively we may assume the magnetization lying in
plane $x-y$ (in such a case the $\overrightarrow{n}$ vector must be along $z-
$axis). \FRAME{ftbpFU}{1.8991in}{1.0724in}{0pt}{\Qcb{Geometry of Josephson
junction with BIS metal as a weak link. The exchange field is directed along
the $z$-axis and the $\mathbf{n}$ vector is along $\ y$-axis. The total
length of the weak link is $2L$.}}{\Qlb{Fig}}{shema_3d_weak_link_2d.jpg}{%
\special{language "Scientific Word";type "GRAPHIC";maintain-aspect-ratio
TRUE;display "USEDEF";valid_file "F";width 1.8991in;height 1.0724in;depth
0pt;original-width 11.3126in;original-height 6.3123in;cropleft "0";croptop
"1";cropright "1";cropbottom "0";filename
'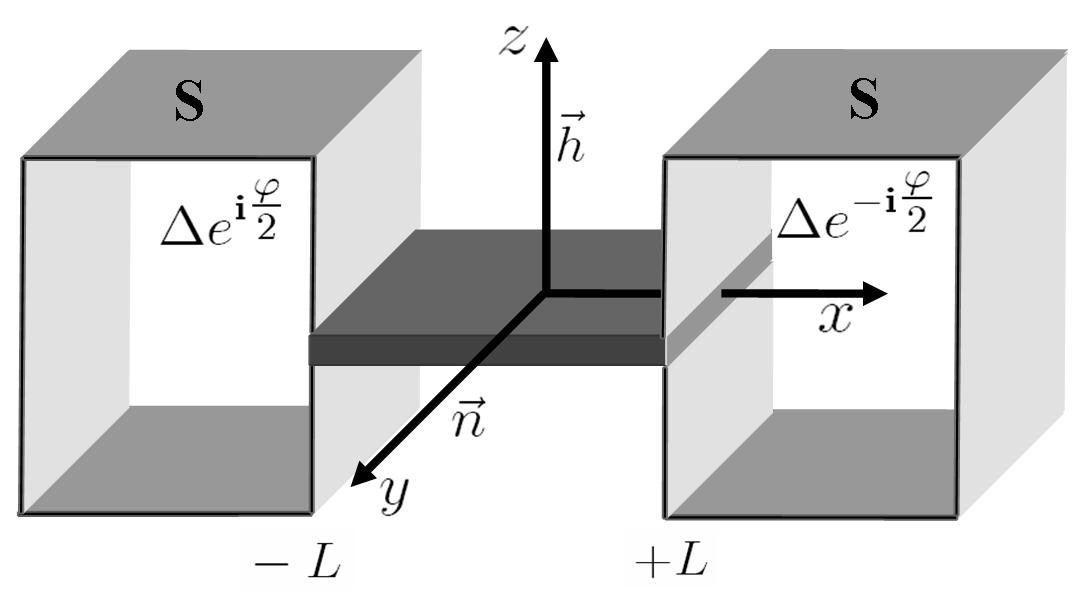';file-properties "XNPEU";}}

In the considered case of the weak link of the length $2L$ , see Fig. 1, the
order parameter depends only on the coordinate $x$ and \ the corresponding
GL equation is

\begin{equation}
a\psi-\gamma\frac{\partial^{2}\psi}{\partial x^{2}}+2i\varepsilon h\frac{%
\partial\psi}{\partial x}=0,  \label{GLequation}
\end{equation}
where we have neglected the non-linear term assuming the superconducting
banks being at temperature slightly below their critical one. The solution
of (\ref{GLequation}) is \ straightforward

\bigskip%
\begin{equation}
\psi=A\exp(q_{1}x)+B\exp(q_{2}x),  \label{exp}
\end{equation}
with $q_{1,2}=i\widetilde{\varepsilon}\pm\sqrt{\frac{a}{\gamma}-\widetilde {%
\varepsilon}^{2}}$, \ where $\widetilde{\varepsilon}=\frac{\varepsilon h}{%
\gamma}$ and the condition $\frac{a}{\gamma}>\widetilde{\varepsilon}^{2}$
assures that the weak link is in the normal state (i. e. the temperature is
above the intrinsic critical temperature of BIS metal $a>a_{c}=\gamma 
\widetilde{\varepsilon}^{2}$). To illustrate the particularity of the
proximity effect in BIS system let us consider the contact of superconductor
with a metal occupying the $x>0$ half-space. In such a case the order
parameter distribution is described by the decaying exponent in (\ref{exp}) $%
\psi\sim\exp(i\widetilde{\varepsilon}x)\exp(-x\sqrt{\frac{a-a_{c}}{\gamma}}%
). $ The difference with the usual proximity effect is that the order
parameter decay is accompanied by the superconducting phase rotation.
Therefore, in the weak link the phase difference proportional to its length
would be accumulated. Due to the $2\pi$ periodicity the actual phase
difference is limited by the interval ($0,2\pi).$

To calculate the current we need to determine the coefficients $A$ and $B$
in (\ref{exp}) from the boundary conditions at the contact with the
superconductors. For illustration we assume that there is no barrier at the
interface and we may use the rigid boundary conditions \cite{GolubovRMP} (i.
e. the normal conductivity of the BIS metal is much smaller than that of the
superconducting bank). Therefore the coefficients $A$ and $B$ are obtained
from the continuity conditions for $\psi$ at $x=\pm L$: $\psi(\pm
L)=\left\vert \Delta\right\vert \exp(\mp i\frac{\varphi}{2})$, with $%
\left\vert \Delta\right\vert $ being the modulus of the order parameter in
the banks and $\varphi$ is the superconducting phase difference across the
junction. Taking into account the new expression for the superconducting
current coming from (\ref{GL}) (note that there is additional contribution
to the current from the spin-orbit term) and performing the corresponding
calculation we readily find in the limit of the long junction $L\sqrt{\frac {%
a}{\gamma}-\widetilde{\varepsilon}^{2}}>>1$

\begin{equation}
j=4e\gamma\left\vert \Delta\right\vert ^{2}\sqrt{\frac{a}{\gamma}-\widetilde{%
\varepsilon}^{2}}\sin(\varphi+2\widetilde{\varepsilon}L)\exp(-2\sqrt{\frac{a%
}{\gamma}-\widetilde{\varepsilon}^{2}}L).  \label{currentGL}
\end{equation}
The current-phase relation 
\begin{equation}
j(\varphi)=j_{c}\sin(\varphi+\varphi_{0})  \label{currentvsphi}
\end{equation}
implies that the junction energy 
\begin{equation}
E_{J}\sim-j_{c}\cos(\varphi+\varphi_{0}),  \label{energy}
\end{equation}
and the minimum energy corresponds to the non-zero phase difference $%
\varphi=-\varphi_{0}$. Naturally the Josephson junction energy may be also
directly obtained from the functional (\ref{GL}). Note that (\ref{energy})
describes the transition from $0$ to $\pi$ junction when $\varphi_{0}$ vary
from $0$ to $-\pi.$ However in contrast to $0$ $-\pi$ transition in
superconductor/ferromagnet/superconductor (S/F/S) junctions \cite{Buzdin-rew}%
, the critical current does not vanish at the transition but remains
constant. The presence of the ground state phase difference $\varphi_{0}$ is
a consequence of BIS and the $h=h_{z\text{ }}$component of the spin field in
the weak link: $\varphi_{0}=\frac{2\varepsilon hL}{\gamma}.$ Note that we
assumed the continuity of the order parameter \ at the bank. In general the
interface barrier provokes a jump of the order parameter and in our case it
produces some additional phase rotation which may even exceed $\varphi_{0}$
for the large values of the interface barrier.

\bigskip The approach on the basis of GL functional thought very insightful
can not adequately describes the systems with strong internal exchange field 
$h>T_{c},$ which is usually the case in the magnetic metals. Therefore we
present also the theory of the $\varphi_{0}$ junction on the basis of the
quasiclassical Eilenberger equations \cite{Eilenberger}.

Provided the spin-orbit interaction is smaller than the characteristic
electron energy scale $E_{F}$ we may consider the Rashba term as and
external potential in the standard scheme of derivation of the Eilenberger
equations (this imply $\alpha<<v$, where $v$ is a Fermi velocity). This
approach has been successfully applied for the description of the
superconducting state in CePt$_{3}$Si \cite{Hayashibis}. In principle to
have a complete description of the BIS junction we need to solve the
Eilenberger equations also in the superconducting banks and take into
account the suppression of the superconducting order parameter near the
interfaces. The full treatment of this problem require the extended
numerical calculations \cite{GolubovRMP}. Below we would like to concentrate
on the peculiar properties of the BIS junctions. That is why we provides the
results for some cases which can be treated analytically. The resulting
coupled equations for anomalous Green function $f_{ij}$ $(\mathbf{v},\mathbf{%
r})$ (matrix in spin space) are rather cumbersome \ for 3D case but strongly
simplified in 2D or 1D case -they are decoupled. Namely for the geometry in
Fig.1 and supposing the weak link to be quasi 2D (in $x-y$ plane) the
Eilenberger equations in clean limit in the region $-L<x<L$ read

\begin{align}
\left( \omega+\frac{v_{x}}{2}\frac{\partial}{\partial x}\right)
f_{12}+\left( ih+\frac{\alpha}{2}\frac{\partial}{\partial x}\right) f_{12} &
=0,  \notag \\
\left( \omega+\frac{v_{x}}{2}\frac{\partial}{\partial x}\right)
f_{21}-\left( ih+\frac{\alpha}{2}\frac{\partial}{\partial x}\right) f_{21} &
=0.  \label{Eilenberg eq}
\end{align}
For the junction Fig.1 the superconducting order parameter in the banks may
be considered constant as a transverse dimension of the BIS metal is small.
Near $T_{c}$ the Eilenberger equation in the bank reads $\left( \omega+%
\frac {v_{x}^{s}}{2}\frac{\partial}{\partial x}\right) f_{12}=\Delta_{R,L}$,
where $v^{s}$ is a Fermi velocity in superconductor and the equation for $%
f_{21}$ is obtained by the substitute $\Delta\rightarrow-\Delta$. From this
equation it follows that for $\omega>0$ and $v_{x}^{s}>0$ the function $%
f_{12}$ is constant in the left bank $f_{12}=$ $\left( \left\vert
\Delta\right\vert /\omega\right) \exp(i\frac{\varphi}{2}),$ while for $%
v_{x}^{s}<0$ it is constant in the right bank $f_{12}=$ $\left( \left\vert
\Delta\right\vert /\omega\right) \exp(i\frac{\varphi}{2})$ \cite{BuzBul82},%
\cite{Baladie}.

Using the corresponding continuity conditions at $x=\pm L$ for the functions 
$f_{ij}$ at the boundary with superconductors we may readily calculate them 
\cite{BuzBul82}. Note that the triplet components of $f_{ij}$ vanish $%
f_{11}=f_{22}=0.$ Knowing the Green functions \ permits readily to calculate
the supercurrent density flowing trough the junction. At temperature close
to the critical temperature $T_{c}$ of the banks at the lowest $\left\vert
\Delta\right\vert ^{2}$ approximation 
\begin{align}
j & =-ieN(0)\pi T_{c}\underset{\omega}{\sum}\left\langle v_{x}\left[ f_{12}(%
\mathbf{v},x)f_{12}^{+}(\mathbf{v},x)\right. \right.  \notag \\
& \left. \left. +f_{21}(\mathbf{v},x)f_{21}^{+}(\mathbf{v},x)\right]
\right\rangle ,  \label{currentEil}
\end{align}
where $N(0)$ is the density of state at the Fermi level. In the limit of the
long junction $L>v/h$ the main contribution in (\ref{currentEil}) comes from
the directions $\left\vert v_{x}\right\vert \lesssim v$ and the formula for
the current takes a very simple form

\begin{equation}
j(\varphi)=j_{0}\sin(\varphi+\frac{4\alpha hL}{v^{2}})\frac{\cos\left( \frac{%
4\left\vert h\right\vert L}{v}+\frac{\pi}{4}\right) }{\sqrt {\frac{%
4\left\vert h\right\vert L}{v}}}.  \label{currentEilof phi}
\end{equation}
Here $j_{0}=eN(0)\frac{v\Delta^{2}}{T_{c}}\left( \frac{\pi}{2}\right) ^{3/2}$
and in the absence of the spin-orbit interaction $\alpha=0$ this expression
coincides with that for $j(\varphi)$ for the 2D S/F/S junction \cite%
{Konschelle}. Comparing (\ref{currentEilof phi}) with $j(\varphi)$ from GL
theory (\ref{currentGL}) we see that the phase shift $\varphi_{0}=\frac{%
4\alpha hL}{v^{2}}$ in both cases is proportional to strength of the
spin-orbit interaction and the product $hL$. On the other hand the critical
current in (\ref{currentEilof phi}) oscillates with $L$ changing its sign.
This is a typical behavior inherent to the S/F/S junctions with the strong
exchange field $h>>T_{c}$ \cite{Buzdin-rew}. Such oscillations are absent in
our GL approach (\ref{currentGL}) as it is adequate for $h\lesssim T_{c}$,
otherwise the gradient terms in [\ref{GL}] changes its sign and it is needed
to introduce the higher derivatives terms. Such modified GL functional
indeed qualitatively describes the oscillatory behavior of the
superconducting order parameter at S/F proximity effect \cite{Buzdin-rew}.

In the 1 D model of the weak link (single channel approximation) the very
similar to (\ref{currentEilof phi}) current-phase dependence is obtained

\begin{equation}
j(\varphi)=eN(0)\frac{\pi v\Delta^{2}}{2T_{c}}\sin(\varphi+\frac{4\alpha hL}{%
v^{2}})\cos\left( \frac{4\left\vert h\right\vert L}{v}\right) .
\label{currentEilof phi1D}
\end{equation}

\bigskip We considered a weak link in the framework of Eilenberger equations
in the clean limit (ballistic regime) . In the diffusive regime \ the very
convenient approach is provided by the Usadel equations \cite{Usadel} for
the Green functions integrated over Fermi surface $F_{ij}(\mathbf{r}%
)=\left\langle f_{ij}(\mathbf{v},\mathbf{r})\right\rangle $. The calculation
of the current on the basis of Usadel approach gives us also the expression
which may be presented in the form (\ref{currentvsphi}). Therefore the
formation of the $\varphi_{0}$ junction by \ the BIS magnets is very general
phenomenon which may be observed in both clean or dirty limits.\bigskip

To summarize in all approaches we obtain the Josephson junction with unusual
current-phase relations $j(\varphi)=j_{c}\sin(\varphi+\varphi_{0}),$ where $%
\ $the phase shift $\varphi_{0}$ is determined by z-component of the
internal magnetic field. Thought our model is applied for the weak
spin-orbit interaction $\alpha<<v$ we may expect that qualitatively the
phase-shift effect would be the same for the systems with strong spin-orbit
interaction $\alpha\sim v$. In this case the characteristic length of the
phase-shift is the same as for the $0-\pi$ transition in S/F/S junctions, i.
e. several $nm$. Therefore we may believe that the formation of $\varphi_{0}$
Josephson junction is inherent to all weak links/barriers with magnetic BIS
metals.

\bigskip The S/F/S Josephson junction may have zero or $\pi$ phase
difference in the ground state depending on the length of the weak link. In
contrast in the $\varphi_{0}$ junction the ground state is always different
from zero and $\pi$ states (except the occasional events $\varphi_{0}$ $=\pi
n$). In the superconducting ring with $\pi$ \ junction the spontaneous
current appears \cite{Buletal} if the parameter $k=\frac{c\Phi_{0}}{2\pi
Lj_{c}}<1$, here $L$ is the inductance of the system. For the $\varphi_{0}$
junction the system energy: 
\begin{equation}
E(\varphi)=\frac{j_{c}}{2e}\left( -\cos(\varphi+\varphi_{0})+\frac {%
k\varphi^{2}}{2}\right) ,
\end{equation}

Therefore the minimum energy is achieved for the phase difference satisfying
the equation

\bigskip%
\begin{equation}
\sin(\varphi+\varphi_{0})+k\varphi=0,
\end{equation}

which always have a non zero solution and then the $\varphi_{0}$ junction
will always generate the spontaneous current with the flux $%
\Phi=-\Phi_{0}\left( \frac{\varphi_{0}}{2\pi}\right) (1-k)$ in $k<<1$ limit$%
. $ The SQUID with one normal and another $\varphi_{0}$ junction would
reveal the shift of the diffraction pattern by $\varphi_{0}$. Note also that
the $\varphi_{0}$ Josephson junctions may serves as a natural phase shifter
in the superconducting electronics circuits.

The very important property of the discussed $\varphi_{0}$ junction is that
it provides a direct mechanism of the coupling between supercurrent and
magnetic moment - indeed the phase shift $\varphi_{0}$ is proportional to $z$
component of the spin field. This means that the precessing magnetization
will be directly coupled with the current which opens new interesting
perspectives to study the coupled magnetic and current dynamics in Josephson
junctions. Applying the voltage to the $\varphi_{0}$ junction we obtain the
Josephson generation and the magnetic moment of the weak link will
experience the effective field varying with Josephson frequency. If this
frequency is close to the ferromagnetic resonance frequency, it may be an
efficient way to generate the spin precessing. Inversely the spin precessing
in the weak link would generate superconducting current in the circuit with $%
\varphi_{0}$ junction.

Finally we note that even in the centosymmetric compounds the inversion
symmetry is broken near the surface. This means that locally the Rashba type
interaction will be present their and then the Josephson junction made by
two superconducting electrodes attached to the surface of ferromagnetic
metal \ would be a $\varphi_{0}$ junction.

The author is grateful to M. Houzet, E. Goldobin, J. Cayssol, F. Konschelle
for useful discussions and comments. This work was supported by French ANR
project "ELEC-EPR".


\bigskip

\begin{thebibliography}{99}
\bibitem{GolubovRMP} A. A. Golubov, M. Yu. Kupriyanov, and E. Il'ichev, Rev.
Mod. Phys. \textbf{76}, 411 (2004).

\bibitem{Josephson} B. D. Josephson, Phys. Lett. \textbf{1}, 251 (1962).

\bibitem{Geshkenbein} V.\ B.\ Geshkenbein, and A. I. Larkin,\ Pis'ma Zh.
Eksp. Teor. Phys. \textbf{43}, 306 (1986), [JETP Lett. \textbf{43}, 395
(1986)].

\bibitem{Yip} S. Yip, Phys. Rev. B \textbf{52}, 3087 (1995).

\bibitem{Sigrist} M. Sigrist, Prog. Theor. Phys. \textbf{99}, 899 (1998).

\bibitem{Tanaka} S. Kashiwaya, and Y. Tanaka, Rep. Prog. Phys. \textbf{63},
1641 (2000).

\bibitem{Goldobin} E. Goldobin et al., Phys. Rev. B \textbf{76}, 224523
(2007).

\bibitem{Bauer} E. Bauer et al., Phys. Rev. Lett. \textbf{92}, 027003 (2004).

\bibitem{Fujimoto} S. Fujimoto, J. Phys. Soc. Jap. \textbf{76}, 051008
(2007).

\bibitem{Samokhin1} K. V. Samokhin, Phys. Rev. B \textbf{70}, 104521 (2004).

\bibitem{Samokhin2} K. V. Samokhin, E. S. Zijlstra, and S. K. Bose, Phys.
Rev. B \textbf{69}, 094514 (2004).

\bibitem{Samokhin3} K. V. Samokhin, Phys. Rev. Lett. \textbf{94}, 027004
(2005).

\bibitem{Kaur} R.\ P.\ Kaur, D. F. Agterberg, and M. Sigrist, Phys. Rev.
Lett. \textbf{94}, 137002 (2005).

\bibitem{Frigeri} P. A. Frigeri, D. F. Agterberg, A. Koga, and M. Sigrist,
Phys. Rev. Lett. \textbf{92}, 097001 (2004).

\bibitem{Edelstein} V. M. Edelstein, Phys. Rev. Lett. \textbf{75}, 2004
(1995).

\bibitem{Dimitrova} O. Dimitrova, and M. V. Figelman, Phys. Rev. B \textbf{76%
}, 014522 (2007).

\bibitem{Hayashi} N. Hayashi, Ch. Iniotakis, M. Machida, and M. Sigrist,
Physica C, in press (2008).

\bibitem{Rashba} E. I. Rashba, Fiz. Tverd. Tela (Leningrad) \textbf{2}, 1224
(1960) [Sov. Phys. Solid State \textbf{2}, 1109 (1960)]; Yu. A. Bychkov and
E. I. Rashba, Pis'ma Zh. Eksp. Teor. Phys. \textbf{39}, 66 (1984),\ [JETP
Lett. \textbf{39, }78 (1984)].

\bibitem{Buzdin-rew} A. I. Buzdin, Rev. Mod. Phys., \textbf{77}, 935 (2005).

\bibitem{Eilenberger} G. Eilenberger, Z. Phys. \textbf{214}, 195 (1968).

\bibitem{Hayashibis} N. Hayashi, K. Wakabayashi, P.\ A.\ Frigeri, and M.
Sigrist, Phys. Rev. B \textbf{73}, 024504 (2006).

\bibitem{BuzBul82} A. I. Buzdin, L. N. Bulaevskii, and S. V. Panyukov, \
Pis'ma Zh. Eksp. Teor. Phys. \textbf{35}, 147 (1982), [JETP Lett. \textbf{35}%
, 178 (1982)].

\bibitem{Baladie} I. Baladi\'{e}, and A. Buzdin, Phys. Rev. B \textbf{64},
224514 (2001).

\bibitem{Konschelle} F. Konschelle, J. Cayssol, and A. Buzdin, to be
published.

\bibitem{Usadel} L. Usadel, Phys. Rev. Lett. 95, 507 (1970).

\bibitem{Buletal} L. N. Bulaevskii, V. V. Kuzii, and A. A. Sobyanin, Pis'ma
Zh. Eksp. Teor. Phys. \textbf{25}, 314 (1977), [JETP Lett., \textbf{25},
290(1977)].
\end{thebibliography}
\end{document}